\documentclass{aa}  
\usepackage{caption}
\usepackage{graphicx}
\usepackage{txfonts}
\begin{document} 
\title{No Robust Statistical Evidence for a Population of Water Worlds in a 2025 Sample of Planets Orbiting M Stars}

   \author{Silke Dainese
          \inst{1}
          \and
          Simon H.\ Albrecht\inst{2}\fnmsep
          }

   \institute{Department of Physics and Astronomy, Aarhus University, Ny Munkegade 120, DK-8000 Aarhus C, Denmark\\
              \email{dainese@phys.au.dk}
         \and
             Department of Physics and Astronomy, Aarhus University, Ny Munkegade 120, DK-8000 Aarhus C, Denmark \\
             \email{albrecht@phys.au.dk}
             }

   \date{Received November 7, 2024 / Accepted February 28, 2025}

  \abstract
   {The study of exoplanets has led to many surprises, one of which is the discovery of planets larger than Earth yet smaller than Neptune, super Earths and gas dwarfs. No such planet is a member of the Solar System, yet they appear to be abundant in the local neighbourhood. Their internal structure is not well understood. Super Earths presumably are rocky planets with a thin secondary atmosphere, whereas gas dwarfs have a substantial (by volume) primary H/He atmosphere. However, conflicting evidence exists regarding the presence of a third class of planets, so-called water worlds, which are hypothesised to contain a significant mass fraction of water in condensed or steam form. This study examines the evidence for water worlds and presents a sample of 60 precisely measured small exoplanets (less than 4 Earth radii) orbiting M dwarf stars. We combine observational data and unsupervised machine-learning techniques to classify these planets based on their mass, radius, and density. We individually model the interior of each planet using the ExoMDN code and classify them into populations based on these models. Our findings indicate that the sample divides into two distinct planet populations, with no clear evidence supporting the existence of water worlds in the current dataset.}

   \keywords{Planets and satellites: fundamental parameters -- Planets and satellites: composition --  Planets and satellites: interiors -- Planets and satellites: terrestrial planets}

   \titlerunning{No robust evidence for a population of water worlds in a 2025 sample}
   \maketitle

\section{Introduction} \label{sec:intro}
Gas dwarfs, planets smaller than Neptune with substantial H/He envelopes and radii between 1.5 and 4 Earth radii, represent a prevalent population of exoplanets in the stellar neighbourhood, standing in stark contrast to their absence within our solar system \citep{Borucki2010, Howard2012, Rogers2015, Fulton+2017}. Within this size range, another hypothesised population of planets is found, often called "water worlds," which are theorised to harbour significant quantities of water in condensed or steam form \citep{Zeng2019, Unterborn2018, Luque2022}.

The potential presence of water worlds gained traction with the study by \cite{Luque2022} who analysed a population of 34 small planets orbiting M dwarfs. Using composition models from \cite{Zeng2019}, they identified six candidates that could possess substantial condensed water layers.
However, subsequent work by \cite{Rogers2023} introduced a layer of complexity by suggesting that the observed characteristics of these planets could instead be explained by atmospheric boil-off processes. Their models propose that planets initially endowed with H/He atmospheres may have shed their outer layers, ultimately becoming stripped rocky cores. Rogers et al. argue that current mass and radius measurements alone are insufficient to establish the presence of a water world population around M dwarfs.
Further insights were provided by \cite{Chakrabarty2023}, who simulated planetary system evolution using formation and atmospheric escape models to reproduce the observed radius valley, a gap between rocky planets and gas dwarfs in the mass-radius diagram. Their findings indicated that 20-35\% of M-dwarf planets without primordial H/He atmospheres could be water-ice-rich, a fraction consistent with the upper limit proposed by \cite{Luque2022}.

Building upon this work, \cite{Parviainen+2024} reanalysed an updated version of the dataset used in\cite{Luque2022} and found no statistical support for a distinct population of water worlds. Their results highlighted the sensitivity of water world classification to the choice of theoretical density models, emphasising that the current data sample is insufficient for definitive conclusions. They also stressed the need for a standardised definition of water worlds to enable meaningful comparisons across studies. 
In addition to these efforts, \cite{Parc2024} conducted a comprehensive analysis using the updated PlanetS catalogue of transiting planets. Their findings aligned with those of \cite{Parviainen+2024}, concluding that no distinct population of water worlds exists around M dwarfs. However, they observed a trend where the minimum mass of gas dwarfs increases with earlier spectral types across the M to F star range, providing important insights into planetary formation processes across different stellar environments. These findings further reinforce the notion that the classification of water worlds remains uncertain and highly model-dependent.

In this work, we adopt a different approach by not relying on a specific compositional definition of water worlds. Instead, we aim to identify an isolated cluster of planets distinct from both the rocky planet and gas dwarf populations, seeking robust statistical evidence through data-driven classification techniques. We investigate the existence of a water world population within an updated sample of small planets by analysing their mass, radius, and density measurements using three different methods.

In section \ref{sec:2024Sample}, we describe our sample and selection criteria. In section \ref{sec:cluster}, we apply unsupervised clustering techniques to the mass-density diagram to identify the number of distinct planetary populations and delineate their boundaries.
In section \ref{sec:ExoMDN}, we apply ExoMDN, a machine-learning-based interior inference model developed by \cite{Baumeister2023}. This model allows us to perform a classification based on the inferred internal structure of the planets, considering physical layers such as the core, mantle, water, and atmosphere, rather than relying solely on mass-radius or mass-density measurements. By incorporating the posterior distributions of these physical layers, our classification leverages the underlying physics encoded within the ExoMDN model. Finally, in section \ref{sec:SynCon}, we discuss our findings and summarise the key conclusions of our analysis.

\section{The 2025 Sample} 
\label{sec:2024Sample}

\begin{figure}[]
    \includegraphics[width=1\columnwidth]{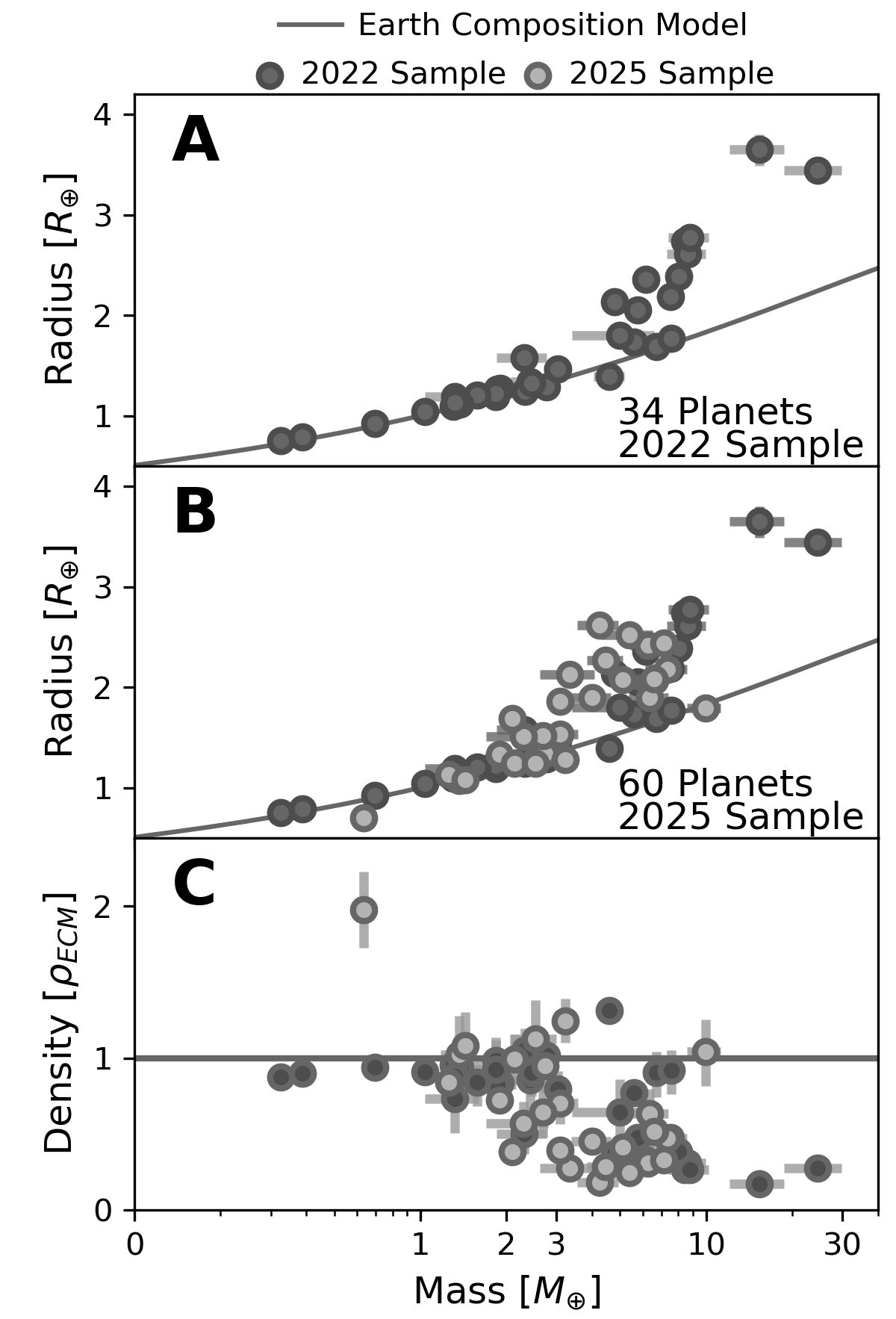}
    \caption{2025 sample. Panel A shows the 2022 sample's planets plotted by mass and radius, while Panel B displays the 2025 sample, with new planets added in 2025 marked in light grey and planets also present in the 2022 sample marked in dark grey. Panel C illustrates the masses and densities of the 2025 sample, comparing the planetary bulk densities to an Earth-like composition model based on \cite{Zeng2019}.}
    \label{fig:Sample_all}
\end{figure}

\cite{Luque2022} compiled a dataset of small planets around M-type host stars, which we here refer to as the 2022 sample.\footnote{L 168-9 b is included in the 2022 sample but excluded from the 2025 sample in this work. This exclusion is due to the planetary parameters reported by \cite{Hobson2024}, which differ from those in \cite{Astudillo-Defru2020} (used in \cite{Luque2022}), having uncertainties in the planet radius that exceed the criteria set for inclusion in this study.}
We started with systems available on the NASA Exoplanet Archive as of January 8th, 2025 and applied similar selection criteria as \cite{Luque2022}. See table~\ref{tabel:2024sample} in the Appendix where the planets new to the 2025 sample are marked with an asterisk. The selection criteria for the two samples are listed in table~\ref{tabel:compare}. When density values were not available in the listed references, we calculated them along with their uncertainties using the equations provided in Appendix \ref{app:density}.

Figure \ref{fig:Sample_all} presents the samples in three panels. Panel A illustrates the masses and radii of planets in the 2022 sample, while panel B shows the masses and radii of planets in the 2025 sample. Panel C displays the masses and densities of the 2025 sample planets, comparing their bulk densities to an Earth-like composition model derived from \cite{Zeng2019}.\footnote{This composition model has been made available at\hspace{0.2em}\url{https://lweb.cfa.harvard.edu/~lzeng/planetmodels.html\#mrrelation} by \cite{Zeng2019}.} All planets below the grey line would have a lower density than a scaled Earth would have, given the model. The model by \cite{Zeng2019} assumes mass fractions of 32.5\% iron and 67.5\% silicates. In this work this model will be referred to as the Earth composition model, ECM for short, and the scaling of density by this model will be referred to as the Earth composition model density, $\rho_{ECM}$.

\begin{table}
\caption{Summary of the selection criteria of the 2022 and 2025 samples.}
\label{tabel:compare}
\centering 
\begin{tabular}{l c c}
\hline\hline 
{Parameter}  & {2022 Sample}  & {2025 Sample} \\
\hline
Data Comprised & 21, July 2021 & 9, January 2025 \\
Host Stars Type & M type & - \\
Max Stellar $T_{Eff}$ & - & 4000K \\
Number of Planets & 34 Planets & 60 Planets  \\ 
Max Radius & $ \textless 4R_{\oplus} $ & $ \textless 4R_{\oplus}$  \\
Radius uncertainty & \textless 8\% & \textless 8\%  \\
Mass uncertainty & \textless 25\% & \textless 25\% \\
Host star radius uncertainty & \textless 5\% & - \\ 
Host star mass uncertainty & \textless 5\% & - \\ 
Star Heliocentric Distance & \textless 100pc & - \\
\hline
\end{tabular}
\end{table}

\section{Population Clustering in mass-density diagram} \label{sec:cluster}
In this section, we sort the 2025 sample into populations based on mass and density. Our approach aims to determine whether the sample is best described by two or three planet populations. This work defines a planet population as a group of planets sharing
similar characteristics in a specific parameter space, such as mass-density.

\subsection{Gaussian Mixture Modelling} \label{sec:GMM}
\begin{figure}[]
    \includegraphics[width=1\columnwidth]{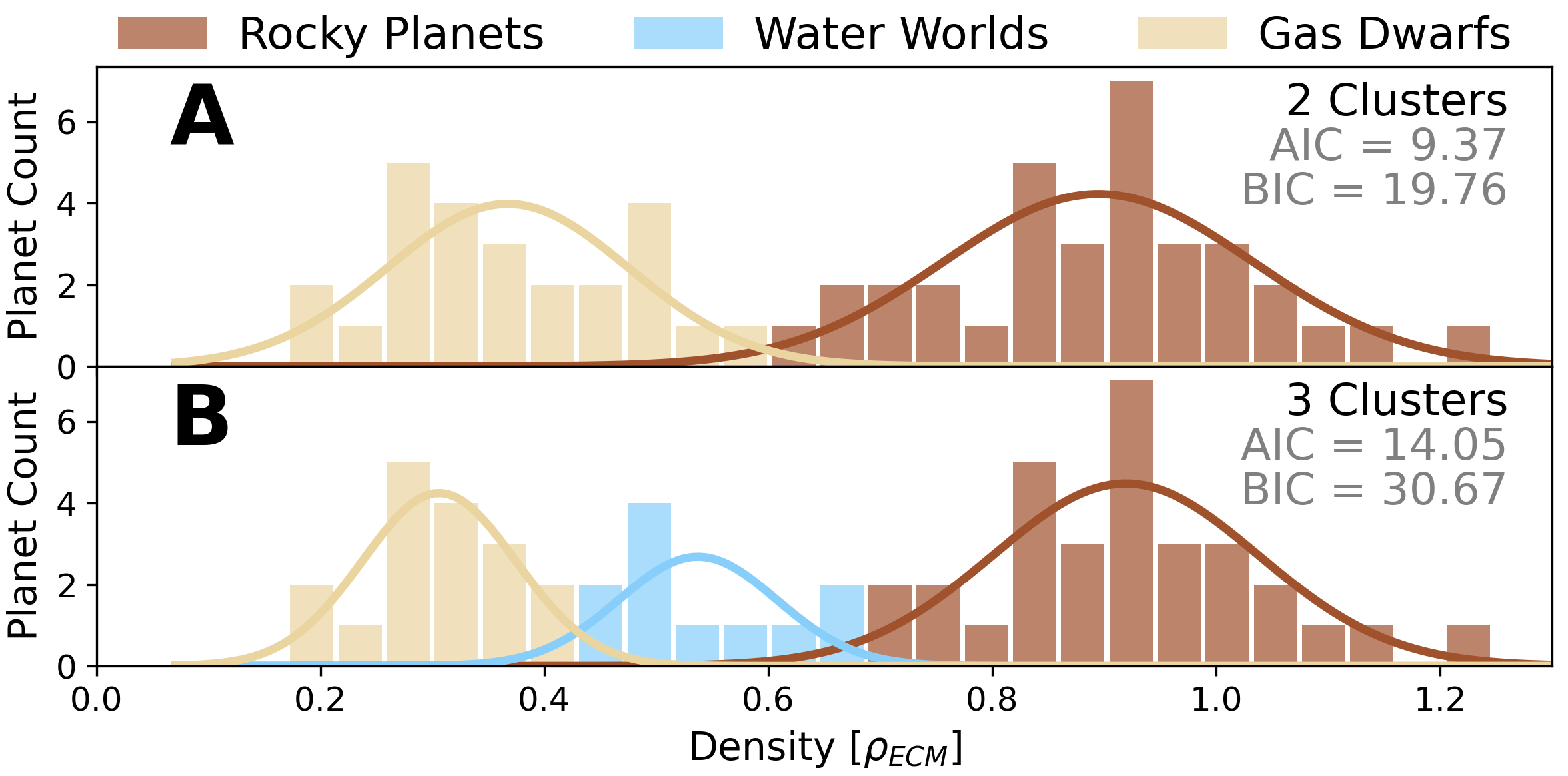}
    \caption{Gaussian Mixture Modelling. The 2025 sample is analysed using Gaussian Mixture Modelling, with planets divided into 21 evenly spaced bins between 0 and 1.5. Panel A assumes two populations, while panel B assumes three populations. Evaluation metrics for the fits are displayed in the top-right corner. GJ 367b is excluded from the analysis due to its high density.}
    \label{fig:GMM_grid}
\end{figure}

We start by applying Gaussian Mixture Models (GMM), a probabilistic machine learning approach that assumes the planets originate from a mixture of a finite number of Gaussian distributions, each characterised by distinct means and variance  \citep{Dempster1977, XU1996}. 
For this purpose, we use the GMM implementation available in the \texttt{scikit-learn} Python package \citep{scikit-learn}\footnote{Documentation:\hspace{0.2em}\url{ https://scikit-learn.org/stable/modules/mixture}.} This clustering algorithm is applied to group the planets based on their  $\rho_{ECM}$ values by effectively fitting Gaussian distributions to the sample histogram. Using GMM enables us to categorise the 2025 sample without requiring prior labelling. Instead, we specify the number of populations, allowing the algorithm to fit the corresponding number of Gaussian components to the dataset.
To determine appropriate initialization parameters, we first apply the K-Means algorithm to set suitable initial means and precisions. K-Means is an unsupervised clustering algorithm that assigns data points to one of K clusters based on their distance to the cluster centres (centroids) \citep{Lloyd1982, macqueen1967}. While the GMM itself does not initialize randomly, its initialization step contains inherent randomness due to the K-Means algorithm. To ensure we identify the best-performing model, we perform 100 initializations during model training. GMM does not inherently take parameter uncertainties into account. To address this, we have implemented a Monte Carlo sampling step during training. In this step, we create a new sample where each planet is drawn from a normal distribution based on its uncertainties (the mean of the upper and lower uncertainty) 100 times. These generated samples are then used to train the algorithm, allowing it to account for uncertainties and improve the robustness of the clustering results.

We apply the GMM algorithm twice: once assuming the sample consists of two populations and again assuming it consists of three populations. The results for the two-population and three-population cases are shown in panel A and B of figure~\ref{fig:GMM_grid}, respectively. \footnote{We exclude GJ 367 b in both GMM runs. It has an extremely high density, $10.2\pm1.3$~g\,cm$^\circ$, \citep{Goffo2023} for its size ($0.699\pm0.024$~R$\oplus$) and therefore clearly does not represent a water world nor a gas dwarf. It would be classified as its own population, taking up its own Gaussian and is not of relevance for our discussion here.}

We use two metrics to evaluate the GMMs' performance against each other and whether the 2025 sample is better described by two or three clusters. The first criterion we use is the Akaike information criterion (AIC). It measures the trade-off between the model's goodness of fit and its complexity (the number of clusters). The second metric is the Bayesian information criterion, BIC, which also includes the sample size. The lower the AIC and BIC the more preferred the model is by this metric. The two criteria are described as follows; 

\begin{eqnarray}
    AIC &=& 2p - 2\ln(\hat{L}) \\
    BIC &=& p\ln(n) - 2\ln(\hat{L})
\end{eqnarray}    

Where \textit{n} is the number of planets in the sample, \textit{p} is the number of parameters learned by the model, and \textit{L} is the maximized value of the likelihood function of the model \citep[e.g.,][]{Geron2019}. The clustering analysis indicates that the two-population model performs best according to both the AIC and BIC criteria. Specifically, for the GMM with two clusters, we obtain AIC = 9.37 and BIC = 19.76, whereas for the GMM with three clusters, AIC increases to 14.05 and BIC to 30.67. This means, that the GMM model suggests, that the planetary densities in this sample are best described by two distinct populations rather than three.
 
\begin{figure*}[]
    \includegraphics[width=1.0\textwidth]{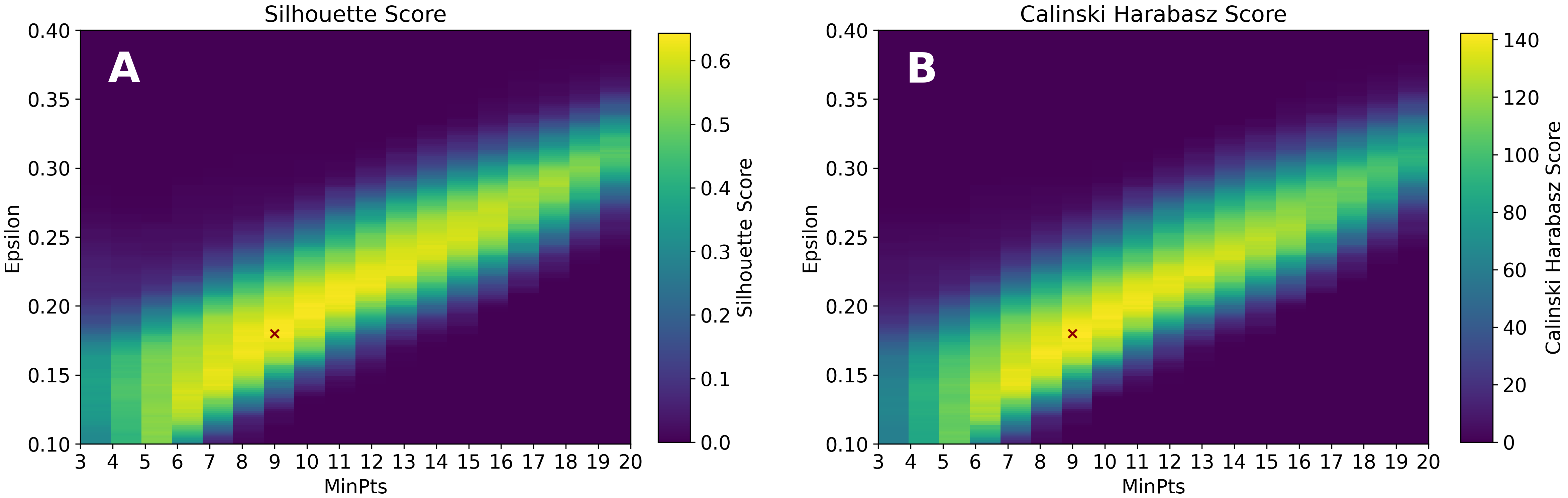}
    \caption{Grids showing the evaluation of the 2025 sample's DBSCAN classification. The evaluation was performed in $\rho_{ECM}$-log10(mass) coordinates with different MinPts and epsilon values. The evaluation is a collection of bootstrapping runs with 200 samples. For each bootstrap sample, two grids of evaluation scores are made by running the DBSCAN on the bootstrapped sample over a range of 200 epsilon values from 0.1 to 0.4 and MinPts from 3 to 20. One grid records the silhouette score, and one grid records the CHI. These two grids are used to find the optimal parameters for the 2025 dataset. In panel A the total grid of all the recorded silhouette scores. In panel B the total grid of all the recorded CHI. In both of the top panels, the best-performing epsilon value is marked with a red x.}
    \label{fig:GridSearch}
\end{figure*}

\subsection{Population Density Based Clustering} \label{sec:DBSCAN}
With GMM our analysis focused solely on $\rho_{ECM}$, requiring us to predefine the number of populations and subsequently compare the results. To overcome this limitation and categorise the 2025 sample into different populations without assuming their number in advance, we employ Density-Based Spatial Clustering of Applications with Noise (DBSCAN) \citep{ester1996, sander1998}. The method is implemented using the \texttt{scikit-learn} Python package\footnote{Documentation:\hspace{0.2em}\url{https://scikit-learn.org/stable/modules/generated/sklearn.cluster.DBSCAN}}. Unlike many other methods, DBSCAN does not assume any particular cluster shape and can handle outliers effectively, i.e., planets not categorised into a population \citep{khan2014,Schubert2017}. We opt to apply DBSCAN to the dataset in the $\log_{10}mass$ versus $\rho_{ECM}$ plane, as this parameter space was used in the analysis by \cite{Luque2022}, and it allows for clustering without the need for physically arbitrary normalisation.
DBSCAN identifies clusters as continuous regions of high density, separated by lower-density regions. DBSCAN has two key parameters, epsilon, and MinPts. The former determines the radius around each data point to assess density, while the latter sets the minimum number of points required to form a dense region.

In order to qualify and record how well the DBSCAN model performs, we use two evaluation metrics; silhouette score \citep{ROUSSEEUW1987} and Calinski–Harabasz index  \citep{Calinski1974}. 

The silhouette score is a measure of how well the planets averagely fit into their own cluster, compared to their nearest neighbour cluster. It is the mean of all planet's silhouette coefficients which is calculated as: 
\begin{equation}
    \textrm{Silhouette Coefficient} = \frac{b - a}{max(a,b)}\,\,\, .
\end{equation}
Here \textit{a} represents the mean distance to particular other planet in the same cluster (giving a measure of cohesion) and \textit{b} is the mean distance to the centre of the nearest neighbouring cluster (giving a measure of separation). The $max$ function simply picks the largest value between $a$ and $b$. The silhouette coefficient ranges from -1 to +1. A value near +1 indicates that a planet is well within its assigned cluster and far from others, whereas a value near 0 suggests proximity to a cluster boundary. A negative value implies potential misclassification. The silhouette score is a mean of all silhouette coefficients in the sample and can thereby also vary between -1 and +1, with larger silhouette scores representing a more successful cluster representation of the sample \citep{Geron2019}. 

The second evaluation metric, the Calinski–Harabasz index (CH index) is also used to evaluate how well the planets are grouped within each cluster and how distinct the clusters are from one another. It measures the balance between cohesion, which refers to how closely packed the planets are within a cluster, and separation, which reflects how far apart the clusters are.

\begin{eqnarray}
    \textrm{CH-Index} &=& \frac{\textrm{cohesion}}{k-1}  \frac{\textrm{separation}}{n-k}^{-1} \\
    \textrm{cohesion} &=& \sum^{k}_{i = 1} n_{i}\left \| \mathbf{c}_{i}-\mathbf{c} \right \|^2 \\
    \textrm{separation} &=& \sum^{k}_{i = 1}\sum_{x\in C_i}^{n_k} \left \| x - \mathbf{c}_k \right \|^2\,\,\,.
\end{eqnarray}  

In these equations, \textit{k} represents the number of clusters, and \textit{n} is the total number of planets in the sample. The term \textit{c} is the centroid of all planets in the sample. For each specific cluster \textit{i}, the number of planets is $n_i$, the set of planets in the cluster is \textit{$C_i$}, and \textit{x} represents an individual planet within the cluster. The centroid of cluster i is denoted as \textit{$c_i$}. A higher CH index value indicates that the clusters are compact and well separated. However, since this is a relative measure, there is no fixed threshold that defines a good clustering result \citep{Liu2010}.

\begin{figure*}[]
    \includegraphics[width=1.0\textwidth]{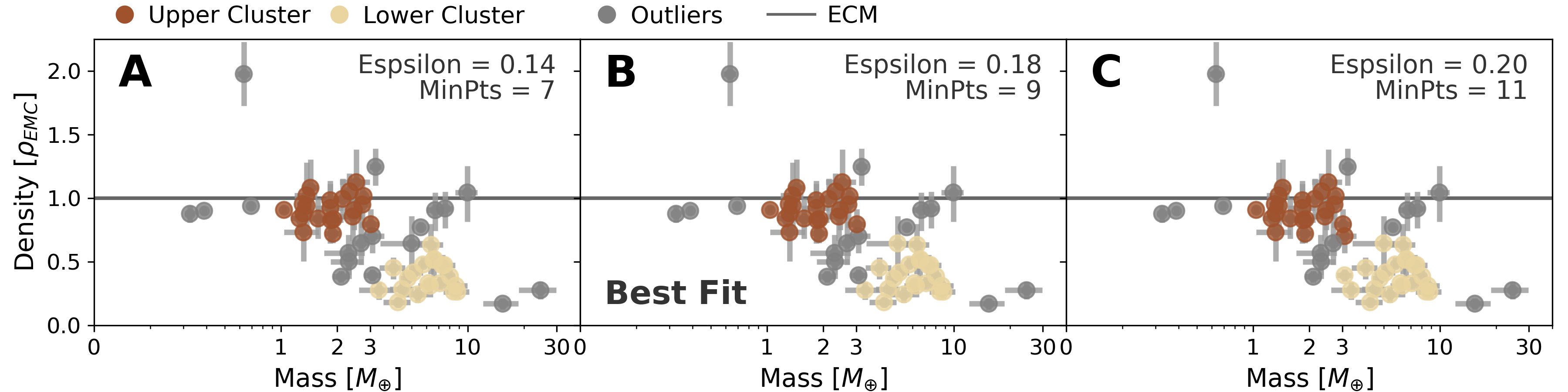}
    \caption{DBSCAN classification of the 2025 sample in $\rho_{ECM}$-log10(Mass) space, using the best-performing combinations different MinPts and epsilon values. Each panel has a different combination of MinPts and epsilon parameters, indicated by the values in the panel. In panel A is seen the fit of the optimal epsilon for MinPts = 7. In panel B, the overall optimal hyperparameters were found from grid search, with MinPts = 9 and epsilon = 0.18 In panel C, the of the optimal epsilon for MinPts = 11. The planets are coloured based on their clusters. Brown planets are rocky planets, and beige planets are gas-rich planets. Grey planets are classified as outliers. The densities are normalized by the Earth composition model (ECM, grey line).}
    \label{fig:DBSCAN_bestfit}
\end{figure*} 

The two parameters, epsilon and MinPts, are set before the learning process begins, and their values influence the learning algorithm's behaviour. Therefore, it is important to find appropriate values for these two parameters before clustering the 2025 sample with DBSCAN. We therefore perform a grid search. Specifically we test 18 values for MinPts from 3 to 20 and 200 values for epsilon from 0.1 to 0.4 resulting in 3600 combinations, only keeping the results that split the sample into at least two populations.\footnote{Fractionally, most of the saved combinations result in two populations. Specifically, 85\% of the saved combinations yield two planet populations, while 10\% produce three populations, 3\% result in four populations, and 2\% lead to five or more populations.} For each combination of MinPts and epsilon, we record both the Silhouette Score and the CH index. To account for the uncertainties in the 2025 sample, we apply bootstrapping, drawing from the uncertainty distributions in mass and density. Specifically, we take 200 draws for each MinPts-epsilon combination, resulting in a total of 720,000 iterations of DBSCAN. The combined results for the Silhouette Scores and the CH index are presented as heatmaps in Figure \ref{fig:GridSearch}. Panel A shows the Silhouette Scores, while Panel B presents the CH index. The combination yielding the highest scores for both metrics is considered optimal. Across both metrics, the best-performing parameters are an epsilon of 0.18 and a MinPts of 9. We have marked this optimal combination with a dark red 'x' in both panels of Figure \ref{fig:GridSearch}.

In figure \ref{fig:DBSCAN_bestfit}, the DBSCAN classification is plotted with three select combinations of the MinPts and epsilon parameters. Each panel corresponds to one combination of MinPts and its best-performing epsilon value. In the figure gas planets are depicted in beige, while rocky planets are shown in brown. Due to the nature of the DBSCAN algorithm, many planets may be identified as outliers, denoted in grey. These outliers are planets that do not fall within high-density regions and are not categorised into the two populations. This does not mean that the outlying planets do not potentially belong to a population; rather, it indicates that they are not situated within the dense regions where most other planets are found on the diagram. The fit with the best performing parameters is seen in panel B of figure \ref{fig:DBSCAN_bestfit}. Here it is seen that the 2025 sample is clustered into two populations, divided by a number of outliers. 

In panel A of figure \ref{fig:DBSCAN_bestfit}, a fit with a MinPts parameter set to 7 is displayed, which is 2 below the optimal parameter found, which is 9. Here, we observe that with lower MinPts and epsilon values, fewer planets are clustered into populations, and more planets are considered outliers. This outcome is expected because when epsilon, the radius around each data point where density is assessed, is smaller, only planets in the highest density areas are included in the clusters. Consequently, fewer planets are classified into populations, resulting in more planets being designated as outliers. However, this clustering also finds two planet populations, not three. In panel C a fit with the MinPts parameter of 11 is displayed. Continuing on the same trend, these panels show that choosing a higher MinPts and epsilon value, increases the number of planets in the clusters, thereby decreasing the number of outliers. 

We find the 2025 sample consistently clustered into two distinct populations with some outliers in between. Using various combinations of epsilon and MinPts around the best-performing values does not change this result. The 2025 sample consistently contains two, not three, clusters with outliers in between.

\begin{figure}[]
    \includegraphics[width=1\columnwidth]{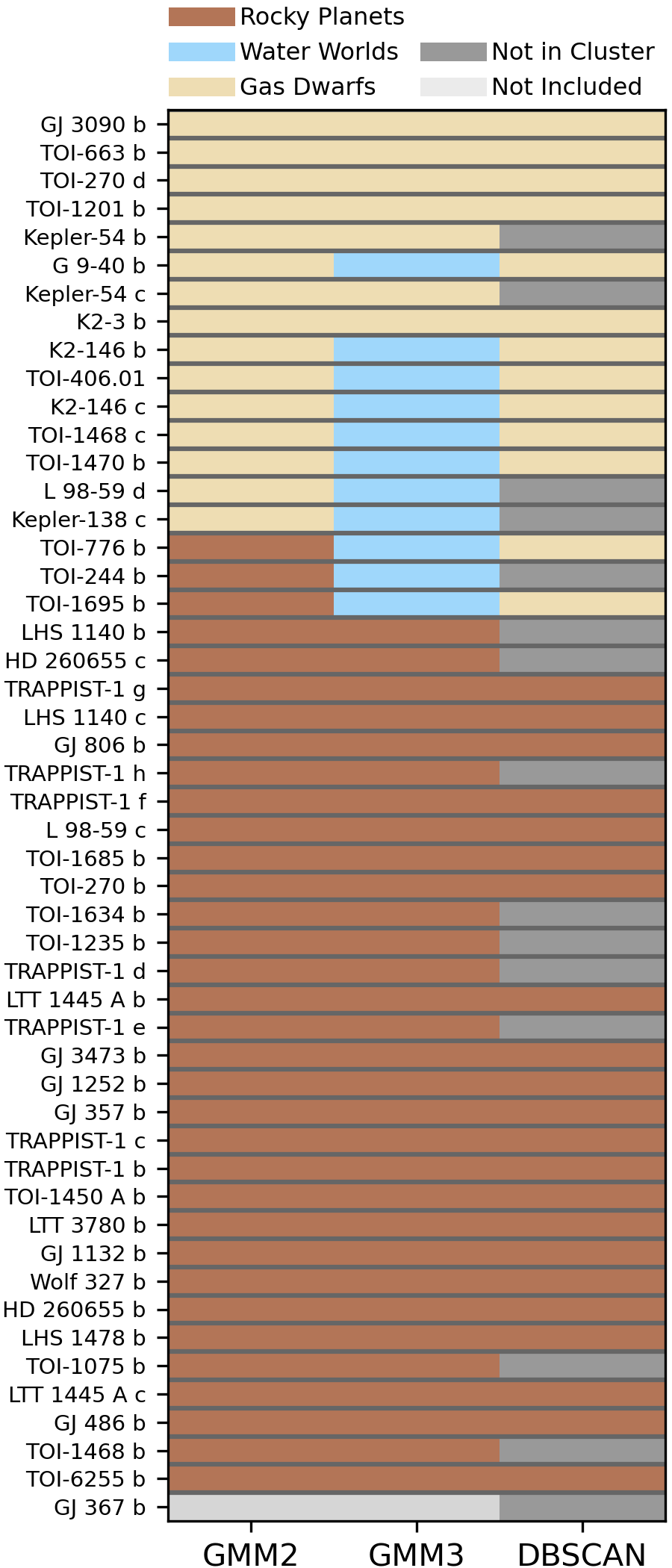}
    \caption{Summary of clustering results. Each row represents a planet, and each column corresponds to a different clustering method. GMM2 is categorization using Gaussian mixture modelling, assuming two populations in the 2025 sample, and GMM3 is categorization using Gaussian mixture modelling, assuming three populations in the 2025 sample. DBSCAN is the population density based clustering using the DBSCAN algorithm on the 2025 sample.}
    \label{fig:Cluster_summery}
\end{figure}

A graphical summary of the results from this section using the different clustering methods is given in figure \ref{fig:Cluster_summery}. Here the 60 planets of the 2025 sample are colour-coded based on the cluster they are assigned to with the different methods. As shown in the figure, TOI-776 b and TOI-1695b are classified as rocky planets using the GMM assuming two populations, but are identified as gas dwarfs by the population density-based clustering. With the exception of these two planets, all other planets classified as either rocky planets or gas dwarfs by the population density-based method receive the same classification when using the two-population GMM. Although the GMM assuming three populations also identifies 11 water worlds, the two-population model is preferred due to its superior performance based on both the AIC and BIC.

\section{Interior Structure-based Clustering} \label{sec:ExoMDN}

\begin{figure*}[]
    \includegraphics[width=1.0\textwidth]{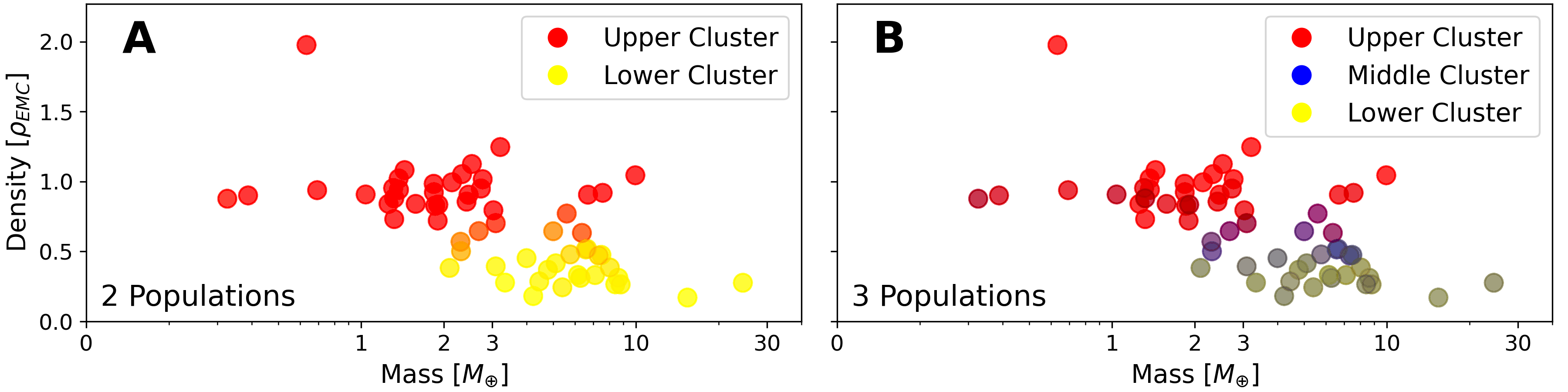}
    \caption{KMeans classification applied to the planet layer thickness fractions derived from ExoMDN for the 2025 sample. A) Classification performed with two assumed populations. B) Classification with three assumed populations. The fitting algorithm was executed 100 times and stacked on top of each other. For each run, a new sample is made by randomly drawing a new composition for each planet from its four-layer posterior distribution, such that the total thickness of a planet adds up to 1. KMeans then clusters each drawn sample, and the planets are coloured according to their assigned cluster. The top cluster (cluster in which the mean density of all member planets is the highest) is coloured red, the bottom cluster is coloured yellow, and the middle (for three populations) is coloured blue. Planets not matching the three primary colours have been assigned to different populations in various iterations.}
    \label{fig:KMeans_Exomdn}
\end{figure*}

So far, our planet clustering has been based on mass, radius, and density. These parameters can be inferred directly from the observations. Here, we incorporate theoretical insights by applying a planetary structure model to infer the interior structure of each of the 60 planets. We will then perform the clustering on the internal structure found for each planet, instead of the mass radius or density measurements for each planet. 

To model the interiors of the planets in the 2025 sample, we use ExoMDN developed by \cite{Baumeister2023}. ExoMDN is an exoplanet interior inference model employing a Mixture Density Network (MDN). This neural network outputs a posterior distribution of mixed Gaussians, allowing us to predict probability distributions for the individual planet's interior structure based on its mass, radius, and equilibrium temperature. ExoMDN is trained using synthetic data generated by the TATOOINE code \cite{Baumeister2020,MacKenzie2023}. TATOOINE divides a planet into four distinct layers: an iron core, a silicate mantle, a water-ice shell, and a hydrogen-helium gas layer. Each layer is characterised by its radius fraction, with the sum of the radii of all layers required to total to the planet's radius. In the future, one might want to incorporate mixing of water into deeper layers of the planet. \cite{Luo2024} proposes the majority of the bulk water can be stored deep within the core and the mantle, and not as only as a layer at the surface, specifically for more massive planets ($>6 M_{\oplus}$) and Earth-sized planets with smaller amounts of water.

Here the core is assumed to consist entirely of molten, pure hcp-iron. The silicate mantle is segmented into an upper and lower part, distinguished by a phase transition occurring at 23 GPa. For the water layers, a composition of pure H$_2$O is assumed, omitting additional compounds like methane or ammonia. The Equations of State (EoS) adopted for modelling span a vast range from 0.1 Pa to 400 TPa and from 150K to $10^{5}$K, encompassing gas, liquid, and solid phases of water \citep{Haldemann2020}. Both liquid and solid layers are considered fully convective, with an adiabatic temperature profile, while water vapor is integrated into the atmosphere. The atmospheric layer is characterised as an isothermal gaseous H/He envelope, mirroring solar-like composition (71\% hydrogen and 29\% helium by mass), based on the EoS from \cite{Saumon1995}, with a temperature equivalent to the equilibrium temperature.

ExoMDN outputs a single multi-dimensional posterior distribution, where each dimension corresponds to the thickness of a layer. Each layer comprises 20 mixture Gaussians each characterised by a mean, spread, and mixing weight.

To categorise the 60 planets based on their four posterior planet layer distributions, we draw a discrete planet composition from each planet's posterior distribution, ensuring that the thickness of each layer—core, mantle, water, and atmosphere—adds up to 1:
\begin{equation}
    1 = X_{Core} + X_{Mantle} +  X_{Water} + X_{Atmosphere}\,\,\, . 
\end{equation}

\begin{figure*}[]
    \includegraphics[width=1.0\textwidth]{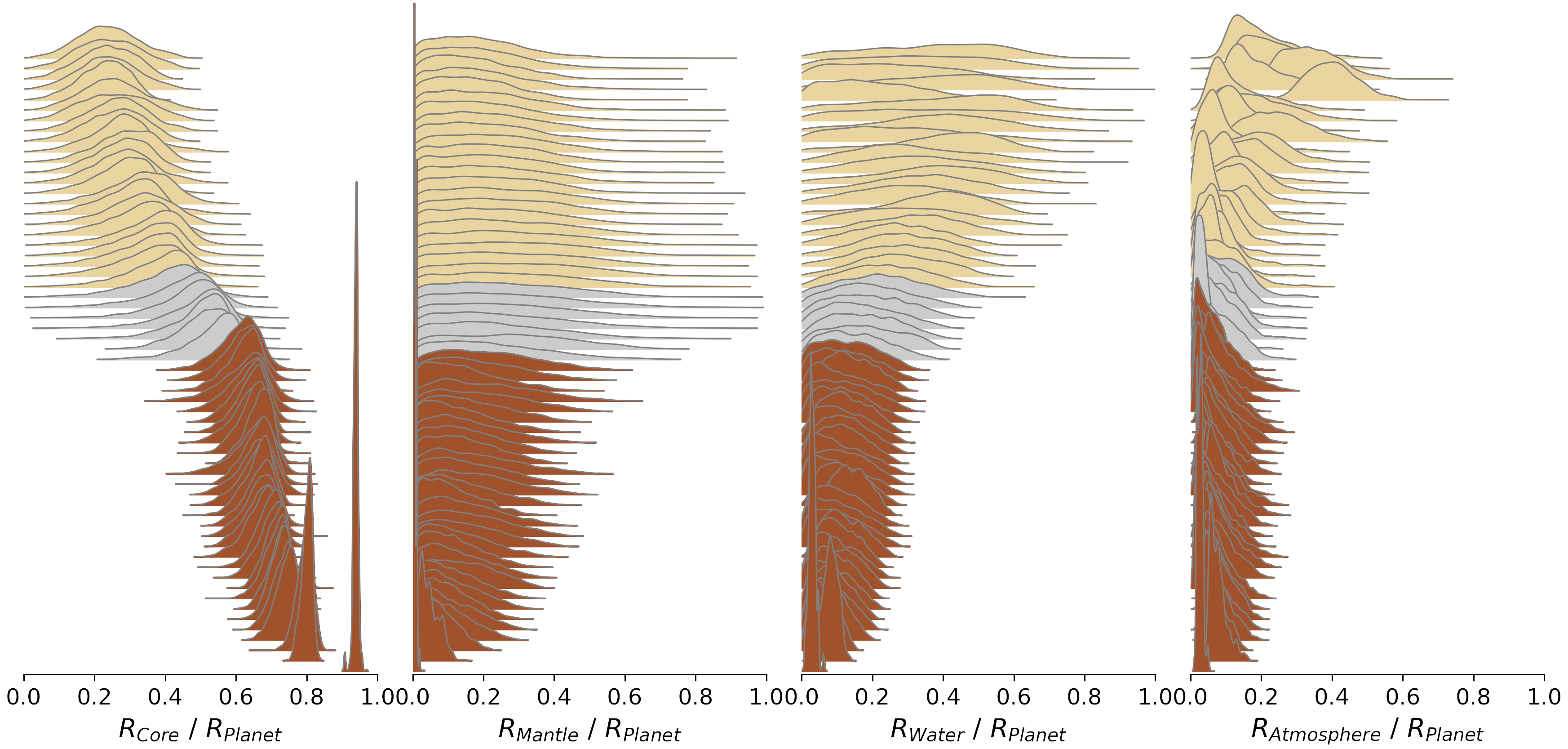}
    \caption{Thickness layer predictions of the planets in the 2025 sample. The planets are sorted top-down by the mode of their predicted core fraction. The plots present the rankings for relative size, core, mantle, water, and atmosphere. The planets are coloured by the KMeans clustering method using two populations, with planets coloured based on their inclusion in the population drawn in at least 90\% of clustering. To see the planet names of each set of posteriors, see the figure~\ref{fig:ExoMDN_all_planets} in the appendix.}
    \label{fig:ExoMDN_prediction}
\end{figure*}

The posterior distributions of the planets in the 2025 sample are shown in figure \ref{fig:ExoMDN_all_planets} in the appendix. Using the predicted thicknesses of each planet, we apply the KMeans\footnote{Documentation:\hspace{0.2em}\url{https://scikit-learn.org/stable/modules/generated/sklearn.cluster.KMeans}} algorithm, implemented in \texttt{scikit-learn}, to group the planets into distinct populations. We run the KMeans algorithm ten times to find the optimal clustering for the four layer thickness predictions. From these, we select the best-performing as the clustering with the lowest measure of the difference between points within the same group, Inertia.

We repeat this procedure 1000 times, i.e., randomly selecting a discrete makeup for each of the 60 planets for each iteration. This iterative process enables us to identify which planets are most challenging to cluster correctly and provides a relative likelihood of a planet being in a particular cluster. We consider a planet part of a population if it is assigned to the same population in 900 out of 1000 iterations. we perform this overall routine twice, assuming two or three populations. 

Figure \ref{fig:KMeans_Exomdn}, displays an overlay of 100 iterations. We assign a colour to each planet group based on the average density of the planets within the group. Specifically, the group with the lowest $\rho_{ECM}$ is coloured yellow, the group with the middle mean $\rho_{ECM}$ is coloured blue, and the group with the highest mean $\rho_{ECM}$ is coloured red.
For the three populations case (panel B), the routine exclusively classifies rocky planets, but cannot distinguish between the lower and middle clusters with sufficient accuracy to classify a single planet into one of these populations. 
The routine performs substantially better for the two population case (panel A), where a population of rocky planets and a population of gas planets are identified. 

In figure \ref{fig:ExoMDN_prediction}, the posterior thickness layer distributions of the 2025 sample are shown. The planets are sorted top-down by the mode of their predicted core fraction. The planets are colour-coded according to their population, as determined by the interior-based two-population K-Means clustering. 

There is a clear distinction between the gas planets at the top and the rocky planets at the bottom. This paper's work so far has been to try to draw these limits and find a division between populations. However, in this figure, no definitive horizontal line separates the two populations. Instead, we observe a gradual evolution in the core fraction of the planets. The top gas planets have small core radius fractions, and the bottom terrestrials are mostly made up of the core. Likewise, most of the large gas planets at the top have extensive atmospheres, while the bottom rocky planets have almost no envelope. The top gas planets also may have an extended water layer, while the bottom terrestrials do not.

\section{Synthesis and Conclusions}  \label{sec:SynCon}
In this work, we have followed up on recent findings by \cite{Luque2022}, who identified three distinct populations among a sample of 34 planets. We extend this work by applying three classification methods to an updated sample that includes 27 additional planets, selected using criteria similar to those in \cite{Luque2022}.

We employed the following three methodologies to investigate the possible presence of a distinct water world population in the 2025 sample:
\begin{enumerate}
    \item Gaussian Mixture Modelling (GMM): This method, detailed in section \ref{sec:GMM}, assumes that planets are divided into two or three populations, each distributed around a mean density normalized by the Earth Composition Model ($\rho_{ECM}$) and scattered in a Gaussian manner. GMM favours the existence of two populations, evaluated by the AIC and BIC. However, it relies solely on the $\rho_{ECM}$ parameter, which has been questioned by \cite{Rogers2023} for its applicability in this context. 
    \item Population Density-Based Clustering (DBSCAN):  Introduced in section \ref{sec:DBSCAN}, this approach allows for flexible population identification in the mass-density plane without assuming a particular shape or requiring the number of populations as an input. Unlike GMM, DBSCAN incorporates the planets' mass uncertainties using a bootstrapping grid search to optimize clustering parameters. The method robustly predicts the presence of two populations with a number of outliers, including L 98-59 d, Kepler-138 c and TOI-244b, which all are categorised as water worlds by the GMM assuming three populations. However, its reliance on normalisation choices and sensitivity to density variations within populations pose limitations. With an optimal MinPts value of 9 in our sample of 60 planets, very small populations, such as a potential water world population, may not be accurately identified. 
    \item Interior Structure-Based Clustering (ExoMDN): In contrast to the other methods, this approach (discussed in section \ref{sec:ExoMDN}) does not rely on $\rho_{ECM}$ but instead uses the ExoMDN machine-learning model \citep{Baumeister2023}, which infers planetary interior compositions from mass, radius and equilibrium temprature. Clustering is performed using KMeans based on the inferred interior layers; core, mantle, water, and atmosphere. When using $K=2$, the results consistently reveal two distinct populations, with outliers that overlap with those identified by DBSCAN.  When $K=3$, the method can only distinguish rocky planets, with the remaining planets frequently switching between the two other clusters. This suggests that the sample lacks sufficient statistical support to confirm the presence of a distinct third population.
\end{enumerate}

Among the three methods, the planets TOI-777 b and TOI-1695 b are classified differently. The population density-based clustering approach identifies them as outliers, whereas DBSCAN classifies them as gas dwarfs, and Gaussian Mixture Modelling categorises them as rocky planets. Overall, both the population density-based clustering and interior structure-based clustering methods yield consistent findings, identifying two dominant populations separated by a set of outliers. The agreement between these independent methods reinforces the conclusion that the planets in this sample do not form three distinct populations.

Our findings align with recent work by \cite{Parviainen+2024}, \cite{Parc2024}, and \cite{Rogers2023}, all of whom report no compelling evidence for a distinct water world population among M-dwarf planets. While \cite{Luque2022} initially suggested the presence of water worlds, our analysis using Gaussian Mixture Modelling, population density-based clustering, and interior structure-based clustering consistently identifies only two populations within the 2025 sample. If water worlds exist, their inferred properties do not form a statistically distinct cluster. This suggests that they either significantly overlap with other planet types or that the current sample lacks the resolution needed to detect them as a separate group. A larger, well-characterised sample, combined with improved interior modelling techniques, will be essential to further investigate the potential existence of water worlds.

\begin{acknowledgements} 
We thank Philipp Bauenmeister and Rafael Luque for their thoughtful comments and discussions, which improved this manuscript. We would also like to thank the anonymous referee for providing constructive and detailed feedback on the manuscript, which greatly enhanced its quality.
We acknowledge the support from the Danish Council for Independent Research through a grant, No.2032-00230B. This research has made use of the NASA Exoplanet Archive, which is operated by the California Institute of Technology, under contract with the National Aeronautics and Space Administration under the Exoplanet Exploration Program.  This work made use of the Python packages \texttt{scikit-learn} \citep{scikit-learn}, and \texttt{ExoMDN} \citep{Baumeister2023}.
\end{acknowledgements}

\bibliographystyle{aa} 
\bibliography{sample.bib}
\appendix
\onecolumn

\section{Sample Data}
\begin{longtable}{llllllcl} 
\caption{Planets in the 2025 sample. Planets new to the 2025 sample and thus not in the 2022 sample are marked with an asterisk.}\label{tabel:2024sample} \\

\hline
Planet name & $\rm Mass_{p}$ & $\rm Radius_{p}$ & $\rm Density_{p}$ & $\rm T_{eq,p}$ & $\rm T_{eff,\star}$ & Primary & Notes \\ 
{} & ($M_{\oplus}$) & {($R_{\oplus}$)} & ($\rho_{\oplus}$) & {(K)} & {(K)} & Reference & {} \\
\hline
\endfirsthead

\caption[]{continued.} \\
\hline
Planet name &  $\rm Mass_{p}$ & $\rm Radius_{p}$ & $\rm Density_{p}$ & $\rm T_{eq,p}$ & $\rm T_{eff,\star}$ & Reference & Notes\\ 
\hline
\endhead

\hline
\endfoot
G 9-40 b\tablefootmark{a}  & $4.0\pm0.63$ & $1.9\pm0.06$ & $3.2^{+0.63}_{0.58}$ & $171\pm1.7$ & $3395\pm51$ & 1 &  \\
GJ 1132 b & $1.84\pm0.19$ & $1.19\pm0.04$ & $5.97^{+0.96}_{-0.79}$ & $584\pm11$ & $3229\pm78$ & 2 &  \\
GJ 1214 b & $8.41^{+0.36}_{-0.35}$ & $2.73\pm0.03$ & $2.26\pm0.11$ & $567\pm8$ & $3101\pm43$ & 3 &  \\
GJ 1252 b & $1.32\pm0.28$ & $1.19\pm0.07$ & $4.2^{+1.3}_{-1.1}$ & $1089\pm69$\tablefootmark{b} & $3458\pm14$ & 4 &\tablefootmark{b} $\rm T_{eq,p}$: 5 \\
GJ 3090 b\tablefootmark{a} & $3.34\pm0.72$ & $2.13\pm0.11$ & $1.89^{+0.52}_{-0.45}$ & $693\pm18$ & $3556\pm70$ & 6 &  \\
GJ 3473 b & $1.86\pm0.3$ & $1.26\pm0.05$ & $5.03^{+1.07}_{-0.93}$ & $773\pm16$ & $3347\pm54$ & 7 &  \\
GJ 357 b & $1.84\pm0.31$ & $1.22\pm0.08$ & $5.6^{+1.7}_{-1.3}$ & $525\pm11$ & $3505\pm51$ & 8 &  \\
GJ 367 b\tablefootmark{a} & $0.63\pm0.05$ & $0.7\pm0.02$ & $10.2\pm1.3$ & $1365\pm32$ & $3522\pm70$ & 9 &  \\
GJ 486 b & $2.77^{+0.08}_{-0.07}$ & $1.29^{+0.02}_{-0.01}$ & $6.66^{+0.23}_{-0.29}$ & $696\pm7$ & $3317\pm36$ & 10 &  \\
GJ 806 b\tablefootmark{a} & $1.9\pm0.17$ & $1.33\pm0.02$ & $4.4\pm0.45$ & $940\pm10$ & $3600\pm16$ & 11 &  \\
HD 260655 b\tablefootmark{a} & $2.14\pm0.34$ & $1.24\pm0.02$ & $6.2\pm1.0$ & $709\pm4$ & $3803\pm10$ & 12 &  \\
HD 260655 c\tablefootmark{a} & $3.09\pm0.48$ & $1.53\pm0.05$ & $4.7^{+0.9}_{-0.8}$ & $557\pm3$ & $3803\pm10$ & 12 &  \\
K2-146 b & $5.77\pm0.18$ & $2.05\pm0.06$ & $3.69\pm0.21$ &$534\pm10$\tablefootmark{b} & $3385\pm70$ & 13 &\tablefootmark{b} $\rm T_{eq,p}$: 14\\
K2-146 c & $7.49\pm0.24$ & $2.19\pm0.07$ & $3.92\pm0.27$ & $520\pm10$ \tablefootmark{b}& $3385\pm70$ & 13 &\tablefootmark{b} $\rm T_{eq,p}$: 15 \\
K2-18 b & $8.63\pm1.35$ & $2.61\pm0.09$ & $2.67^{+0.52}_{-0.47}$ & $255\pm4$ & $3457\pm39$ & 16 &  \\
K2-25 b & $24.5^{+5.7}_{-5.2}$ & $3.44\pm0.12$ & $3.31^{+0.84}_{-0.78}$ \tablefootmark{c}& $494\pm11$ & $3207\pm58$ & 17 &  \\
K2-3 b\tablefootmark{a} & $5.11^{+0.65}_{-0.64}$ & $2.08^{+0.08}_{.0.07}$ & $3.11^{+0.49}_{-0.46}$ & $501\pm5$ & $3844\pm61$ & 18 &  \\
Kepler-138 c\tablefootmark{a} & $2.3^{+0.6}_{-0.5}$ & $1.51\pm0.04$ & $3.6^{+1.1}_{-0.9}$ & $410\pm8$ & $3841\pm50$ & 19 &  \\
Kepler-54 b\tablefootmark{a} & $3.09\pm0.3$ & $1.86\pm0.06$ & $2.64^{+0.31}_{-0.32}$ & 453\tablefootmark{b} & $3853\pm80$ & 20 &\tablefootmark{b} $\rm T_{eq,p}$: 21 \\
Kepler-54 c\tablefootmark{a} & $2.1\pm0.2$ & $1.69\pm0.05$ & $2.38\pm0.29$ & 395\tablefootmark{b} & $3853\pm80$ & 20 &\tablefootmark{b} $\rm T_{eq,p}$: 21 \\
L 98-59 c & $2.42^{+0.35}_{-0.34}$ & $1.34\pm0.07$ & $5.46^{+1.23}_{-1.05}$ &$553\pm27$ \tablefootmark{b}& $3412\pm49$ & 4 &\tablefootmark{b} $\rm T_{eq,p}$: 22 \\
L 98-59 d & $2.31^{+0.46}_{-0.45}$ & $1.58\pm0.08$ & $3.17^{+0.85}_{-0.73}$ & $416\pm20$ \tablefootmark{b}& $3412\pm49$ & 4 &\tablefootmark{b} $\rm T_{eq,p}$: 22 \\
LHS 1140 b\tablefootmark{a} & $5.6\pm0.19$ & $1.73\pm0.02$ & $5.9\pm0.3$ & $226\pm4$ & $3096\pm48$ & 23 &  \\
LHS 1140 c\tablefootmark{a} & $1.91\pm0.06$ & $1.27\pm0.03$ & $5.1\pm0.4$ & $422\pm7$ & $3096\pm48$ & 23 &  \\
LHS 1478 b & $2.33\pm0.2$ & $1.24\pm0.05$ & $6.67^{+1.03}_{-0.89}$ & $595\pm10$ & $3381\pm54$ & 24 &  \\
LP 791-18 c\tablefootmark{a} & $7.1\pm0.7$ & $2.44\pm0.1$ & $2.69\pm0.41$\tablefootmark{c} & $324\pm2$ & $2960\pm55$ & 25 &  \\
LTT 1445 A b\tablefootmark{a} & $2.73^{+0.25}_{-0.23}$ & $1.34^{+0.11}_{-0.06}$ & $6.2^{+1.2}_{-1.3}$ & $431\pm23$ & $3340\pm150$\tablefootmark{b} & 26 &\tablefootmark{b} $\rm T_{eff,\star}$: 27 \\
LTT 1445 A c\tablefootmark{a} & $1.37\pm0.19$ & $1.07^{+0.1}_{-0.07}$ & $5.9^{+1.8}_{-1.5}$ & $516\pm28$ & $3340\pm150$\tablefootmark{b}& 26 &\tablefootmark{b} $\rm T_{eff,\star}$: 27\\
LTT 3780 b & $2.46\pm0.19$ & $1.32\pm0.06$ & $5.8^{+1.0}_{-0.8}$ & $903\pm26$ & $3358\pm92$ & 28 &  \\
LTT 3780 c & $8.04^{+0.5}_{-0.48}$ & $2.39^{+0.1}_{-0.11}$ & $3.24^{+0.55}_{-0.43}$ & $359\pm10$ & $3358\pm92$ & 28 &  \\
TOI-1075 b\tablefootmark{a} & $9.95^{+1.36}_{-1.3}$ & $1.79^{+0.12}_{-0.08}$ & $9.32^{+2.05}_{-1.85}$ & $1323\pm44$ & $3875\pm75$ & 29 &  \\
TOI-1201 b\tablefootmark{a} & $6.28^{+0.84}_{-0.88}$ & $2.42^{+0.09}_{-0.09}$ & $2.45^{+0.48}_{-0.42}$ & $703\pm15$ & $3476\pm51$ & 30 &  \\
TOI-1231 b & $15.4\pm3.3$ & $3.65^{+0.16}_{-0.15}$ & $1.74^{+0.47}_{-0.42}$ & $330\pm4$ & $3553\pm51$ & 31 &  \\
TOI-1235 b & $6.69^{+0.67}_{-0.69}$ & $1.69\pm0.08$ & $7.25^{+1.3}_{-1.1}$ & $775\pm13$ \tablefootmark{b}& $3997\pm51$ & 4 &\tablefootmark{b} $\rm T_{eq,p}$: 32 \\
TOI-1266 b\tablefootmark{a} & $4.23\pm0.69$ & $2.62\pm0.11$ & $1.3\pm0.3$ & $425\pm20$ & $3618\pm157$ & 33 &  \\
TOI-1450 A b\tablefootmark{a} & $1.26\pm0.13$ & $1.13\pm0.04$ & $4.79\pm0.73$ \tablefootmark{c}& $722\pm35$ & $3437\pm86$ & 34 &  \\
TOI-1468 b\tablefootmark{a} & $3.21\pm0.24$ & $1.28\pm0.04$ & $8.41^{+0.99}_{-0.98}$ \tablefootmark{c}& $682\pm7$ & $3496\pm25$ & 35 &  \\
TOI-1468 c\tablefootmark{a} & $6.64^{+0.67}_{-0.68}$ & $2.06\pm0.04$ & $4.15\pm0.5$ \tablefootmark{c}& $338\pm4$ & $3496\pm25$ & 35 &  \\
TOI-1470 b\tablefootmark{a} & $7.32^{+1.21}_{-1.24}$ & $2.18\pm0.04$ & $3.86^{+0.7}_{-0.68}$ & 734 & $3709\pm11$ & 36 &  \\
TOI-1634 b & $7.57^{+0.71}_{-0.72}$ & $1.77\pm0.08$ & $7.6^{+1.3}_{-1.1}$ & $924\tablefootmark{b}\pm22$ & $3472\pm70$ & 4 &\tablefootmark{b} $\rm T_{eq,p}$: 37\\
TOI-1685 b & $3.03^{+0.33}_{-0.32}$ & $1.47\pm0.05$ & $5.3\pm0.8$ & $1062\pm27$ & $3575\pm53$ & 38 &  \\
TOI-1695 b\tablefootmark{a} & $6.36\pm1.0$ & $1.9^{+0.16}_{-0.14}$ & $5.0^{+1.8}_{-1.3}$ & $698\pm14$ & $3690\pm50$ & 39 &  \\
TOI-244 b\tablefootmark{a} & $2.68\pm0.3$ & $1.52\pm0.12$ & $4.2\pm1.1$ & $458\pm20$ & $3433\pm100$ & 40 &  \\
TOI-269 b & $8.8\pm1.4$ & $2.77\pm0.12$ & $2.28^{+0.48}_{-0.42}$ & $531\pm25$ & $3514\pm70.$ & 41 &  \\
TOI-270 b & $1.58\pm0.26$ & $1.21\pm0.04$ & $4.97\pm0.94$ & $581\pm14$ & $3506\pm70$ & 42 &  \\
TOI-270 c & $6.15\pm0.37$ & $2.36\pm0.06$ & $2.6\pm0.26$ & $488\pm12$ & $3506\pm70$ & 42 &  \\
TOI-270 d & $4.78\pm0.43$ & $2.13\pm0.06$ & $2.72\pm0.33$ & $387\pm10$ & $3506\pm70$ & 42 &  \\
TOI-406.01\tablefootmark{a} & $6.57^{+1.0}_{-0.9}$ & $2.08^{+0.16}_{-0.15}$ & $4.1\pm1.1$ & $368\pm14$ & $3392\pm44$ & 43 &  \\
TOI-4438 b\tablefootmark{a} & $5.4\pm1.1$ & $2.52\pm0.13$ & $1.85^{+0.51}_{-0.44}$ & $435\pm15$ & $3422\pm81$ & 44 &  \\
TOI-6255 b\tablefootmark{a} & $1.44\pm0.14$ & $1.08\pm0.06$ & $6.3\pm1.29$ \tablefootmark{c} & 1256 & $3421\pm70$ & 45 &  \\
TOI-663 b\tablefootmark{a} & $4.45^{+0.63}_{-0.66}$ & $2.27^{+0.1}_{-0.09}$ & $2.07^{+0.42}_{-0.38}$ & $674\pm30$ & $3681\pm70$ & 46 &  \\
TOI-776 b & $5.0\pm1.6$ & $1.8\pm0.08$ & $4.8^{+1.8}_{-1.6}$ & $520\pm12$ & $3725\pm60$ & 47 &  \\
TRAPPIST-1 b & $1.37\pm0.07$ & $1.12\pm0.01$ & $5.44^{+0.26}_{-0.28}$ & $397\pm3.8$\tablefootmark{b} & $2566\pm26$ & 48 &\tablefootmark{b} $\rm T_{eq,p}$: 49 \\
TRAPPIST-1 c & $1.31\pm0.06$ & $1.1\pm0.01$ & $5.46^{+0.22}_{-0.24}$ & $339\pm3.3$\tablefootmark{b} & $2566\pm26$ & 48 &\tablefootmark{b} $\rm T_{eq,p}$: 49 \\
TRAPPIST-1 d & $0.39\pm0.01$ & $0.79\pm0.01$ & $4.37^{+0.15}_{-0.17}$ & $286\pm2.8$\tablefootmark{b}& $2566\pm26$ & 48 &\tablefootmark{b} $\rm T_{eq,p}$: 49 \\
TRAPPIST-1 e & $0.79\pm0.01$ & $0.92\pm0.01$ & $4.9^{+0.17}_{-0.18}$ & $249\pm2.4$\tablefootmark{b}& $2566\pm26$ & 48 &\tablefootmark{b} $\rm T_{eq,p}$: 49\\
TRAPPIST-1 f & $1.04\pm0.03$ & $1.04\pm0.01$ & $5.02^{+0.14}_{-0.16}$ & $217\pm2.1$\tablefootmark{b} & $2566\pm26$ & 48 &\tablefootmark{b} $\rm T_{eq,p}$: 49 \\
TRAPPIST-1 g & $1.32\pm0.04$ & $1.13^{+0.02}_{-0.01}$ & $5.06^{+0.14}_{-0.16}$ & $197\pm1.9$ \tablefootmark{b}& $2566\pm26$ & 48 &\tablefootmark{b} $\rm T_{eq,p}$: 49 \\
TRAPPIST-1 h & $0.33\pm0.02$ & $0.76\pm0.01$ & $4.16\pm0.33$ & $171\pm1.7$ \tablefootmark{b}& $2566\pm26$ & 48 &\tablefootmark{b} $\rm T_{eq,p}$: 49 \\
Wolf 327 b\tablefootmark{a} & $2.53\pm0.46$ & $1.24\pm0.06$ & $7.24\pm1.66$ & $996\pm18$ & $3542\pm70$ & 50 & 
\end{longtable}

\begin{table*}[h]
    \centering
    \caption*{\textbf{Notes.} All planetary and stellar parameters used in this study. $\rm Mass_{p}$, $\rm Radius_{p}$ and $\rm Density_{p}$ are the planet mass, radius and density. $\rm T_{eq,p}$ is the planets equilibrium temperature and $\rm T_{eff,\star}$ is the stellar effective temperature. \\
    \tablefoottext{a}{Planets new to the 2025 sample, that where not in the 2022 sample.}
    \tablefoottext{b}{Planets where the stellar effective temperature or the planet equilibrium temperature was unavailable in the primary reference. The secondary reference is indicated in the notes.} 
    \tablefoottext{c}{Planetary bulk densities which have been calculated in this work using the equations of appendix \ref{app:density}}.  
\\    
    \textbf{References:}
1: \cite{Luque2022G940b},
2: \cite{Xue2024},
3: \cite{Mahajan2024},
4: \cite{Luque2022},
5: \cite{Shporer2020},
6: \cite{Almenara2022},
7: \cite{Kemmer2020},
8: \cite{Luque2019},
9: \cite{Goffo2023},
10: \cite{WeinerMansfield2024},
11: \cite{Palle2023},
12: \cite{Luque2022b},
13: \cite{Hamann2019},
14: \cite{Livingston2018},
15: \cite{Lam2020},
16: \cite{Benneke2019},
17: \cite{Stefansson2020},
18: \cite{Diamond-Lowe2022},
19: \cite{Piaulet2023},
20: \cite{Leleu2023},
21: \cite{Muirhead2012},
22: \cite{Demangeon2021},
23: \cite{Cadieux2024},
24: \cite{Soto2021},
25: \cite{Peterson2023},
26: \cite{Pass2023},
27: \cite{Oddo2023},
28: \cite{Bonfanti2024},
29: \cite{Essack2023},
30: \cite{Kossakowski2021},
31: \cite{Burt2021},
32: \cite{Bluhm2020},
33: \cite{Cloutier2024},
34: \cite{Brady2024},
35: \cite{Chaturvedi2022},
36: \cite{Gonzalez-Alvarez2023},
37: \cite{CloutierTOI-1634b}
38: \cite{Burt2024},
39: \cite{Cherubim2023},
40: \cite{Castro-Gonzalez2023},
41: \cite{Cointepas2021},
42: \cite{Van-Eylen2021},
43: \cite{Lacedelli2024},
44: \cite{Goffo2024},
45: \cite{Dai2024},
46: \cite{Cointepas2024},
47: \cite{Luque2021},
48: \cite{Agol2021},
49: \cite{Ducrot2020},
50: \cite{Murgas2024}.
}
\end{table*}

\section{Density Calculations} \label{app:density}
We have calculated the densities and their uncertainties for planets that do not have listed densities, by using standard uncertainty propagation. In the equations below, we describe the methods used to compute the densities, as well as the respective upper and lower uncertainties.
\begin{eqnarray} 
    \rho &=&  \frac{3M}{4\pi R^{3}}\\ \label{eq:density1}
    (\Delta\rho)^2 &=& \left(\frac{\delta\rho}{\delta M} \Delta M \right)^2 + \left(\frac{\delta\rho}{\delta R} \Delta R\right)^2 = \left(\frac{3}{4\pi R^3} \Delta M \right)^2 + \left(\frac{9M}{4\pi R^4} \Delta R\right)^2 \label{eq:density2}
\end{eqnarray}    
    
For asymmetric (upper and lower) uncertainties, we calculate:
\begin{eqnarray}
        \rho_{+} &=& \sqrt{\left(\frac{3}{4\pi R^3}M_{+} \right)^2 + \left(\frac{9M}{4\pi R^4} R_{-} \right)^2 } \label{eq:density3} \\
        rho_{-} &=& \sqrt{\left(\frac{3}{4\pi R^3}M_{-} \right)^2 + \left(\frac{9M}{4\pi R^4} R_{+} \right)^2 } \label{eq:density4}
\end{eqnarray}
Where:
\begin{itemize}
    \item $M$ is the mass of the planet, and $\Delta M$ is its uncertainty. $M_{+}$ and $M_{-}$ denote the upper and lower uncertainties (errors) in the planet’s mass, respectively.
    \item $R$ is the radius of the planet, and $\Delta R$ is its uncertainty. $R_{+}$ and $R_{-}$ denote the upper and lower uncertainties (errors) in the planet’s radius, respectively.
    \item $\rho$ is the calculated bulk density of the planet, and $\Delta \rho$ represents the (symmetric) uncertainty in the calculated density.
    \item $\frac{\partial \rho}{\partial M}$ and $\frac{\partial \rho}{\partial R}$ denote the partial derivatives of the density with respect to mass and radius, respectively.
\end{itemize}

In summary, equation \ref{eq:density1} gives the density, equation \ref{eq:density2} shows how to propagate symmetric uncertainties using partial derivatives, and equations \ref{eq:density3} and \ref{eq:density4} illustrate how to obtain the upper and lower bounds on the density, when the uncertainties in mass and radius are asymmetric.

\newpage
\section{Posterior Distributions of the 2025 Sample} \label{app:fig}
\begin{figure*}[hb]
  \centering
\includegraphics[height=0.87\textheight,keepaspectratio]{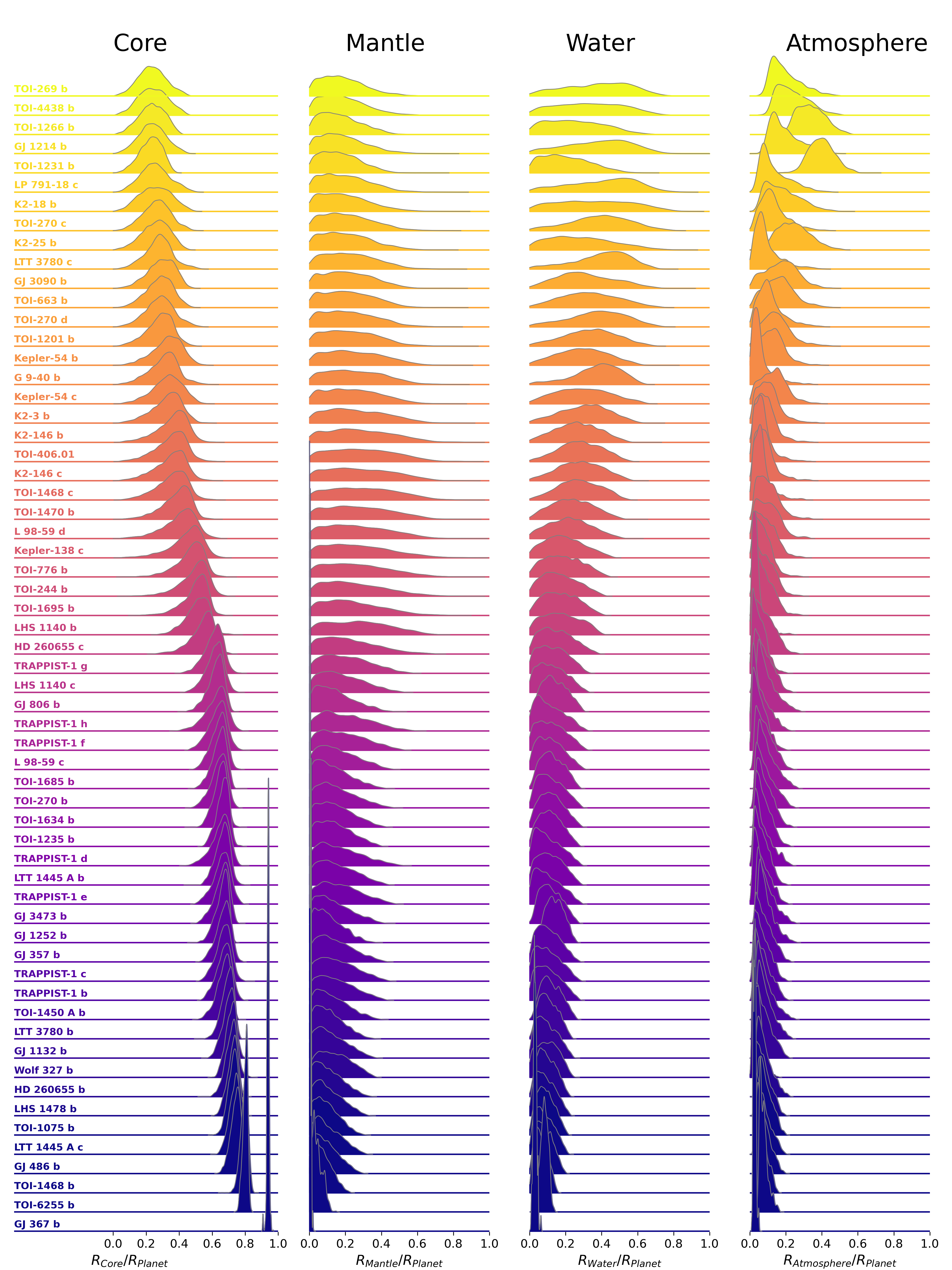}
  \caption{Planet layer predictions for the planets in the 2025 sample. The planets are sorted top-down by the mode of their predicted core fraction.}
  \label{fig:ExoMDN_all_planets}
\end{figure*}

\end{document}